**Great enhancement of mechanical features in PLA based composites containing aligned few layer graphene (FLG), the effect of FLG loading, size and dispersion on mechanical and thermal properties.**

*Hamza El Marouazi,[a] Benoit van der Schueren,[a] Damien Favier,[b] Anaëlle Bolley,[c] Samuel Dagorne,[c] Thierry Dintzer,[a] Izabela Janowska[a*]*

[a] Institut de Chimie et Procédés pour l'Énergie, l'Environnement et la Santé (ICPEES), CNRS UMR 7515-Université de Strasbourg, 25 rue Becquerel 67087 Strasbourg, France

[b] Institute Charles Sadron (ICS), UPR 22- Université de Strasbourg, 23 rue de Loess 67200 Strasbourg, France

[c] Institute de Chimie de Strasbourg, CNRS UMR 7177-Université de Strasbourg, 4 rue Blaise Pascal 67008 Strasbourg, France

Four series of PLA based composite films containing horizontally aligned few layer graphene (FLG) flakes of high aspect ratio and adsorbed albumin are prepared. The mechanical and thermal properties vary with percentage, dispersion degree and size of FLG flakes. Great improvement up to 290 and 360% of tensile modulus and strength respectively was obtained for the composite containing high lateral size of FLG at 0.17% wt., and up to 60 and 80% for the composite with very well dispersed 0.02% wt. FLG. The composites of PLA and PEG-PLLA containing very well dispersed FLG flakes at 0.07 % wt. are ductile showing enhancement of elongation at break up to respectively 80 and 88%. Relatively high electrical conductivity, $5 \times 10^{-3}$ S/cm, is measured for PLA film charged with 3% of FLG.

**1. Introduction**

According to sustainable development, the progress in material science is inseparable from efficiency and environmental issues. This concerns the nature of raw materials, their preparation, as well as their degradation and/or recycling ability. Among the materials addressing environmental issues, bio-polymers with polylactid (PLA) are at the forefront. PLA, a bio-degradable and produced from renewable resources polymer, has already found numerous applications in several domains such as packaging, textile fibers, and especially biomedicine,[1, 2, 3] but some of its properties still require improvements. The enhancement of mechanical and thermal resistance, crystallinity rate or conductivity of PLA are expected with the addition of specific fillers. Among these fillers, nanocarbons and, more recently, 2D graphene-based materials (GBM) receive special attention. [3] GBM are often graphene oxide (GO) and reduced graphene oxide (rGO); [4] graphene platelets (GP) or flakes that are few- or multilayer graphene (FLG/MLG), graphene and functionalized graphene are less investigated. [5-8] The crystallinity/chemistry (degree of reduction and amount of heteroatoms), planar size/thickness and consequently aspect ratio of GBM vary depending on the GBM structure, having an impact on the properties of GBM and next of their composites. Another important aspect affecting the final features of the composites is the dispersion of GBM often linked to their interactions with the polymer. For this purpose, some functionalizations of graphene were investigated. [8-11] Likewise, the way of preparation of GBM and of composites has an additional impact. Several studies reported on PLA composites containg different GBM additives, mostly films, have been reported. Thus, the enhancement of tensile strength and elongation at break by 17% and 51% respectively was achieved for the PLA composites containing 0.1 wt. % of "graphene" obtained from the exfoliated expanded graphite.[5] The effect of different aspect ratio (lateral size vs. thickness) of several graphene platelets fillers in "hard" PLA and "soft" polymer on elastic modulus was investigated. While enhancement of 200% was obtained in the case of casted and hot-pressed "soft" polymer, in PLA a maximum improvement of 35 % and 17% was achieved for large MLG and FLG, respectively, with a loading of 5 wt%. [6] The "average" enhancement in PLA was attributed to the lack of MLG/FLG flakes' orientation, contrary to the soft polymer, where horizontal alignment could be induced via hot-pressing. Y. Gao et al. reported a maximum improvement of Young modulus of 10 and 24% for smaller and larger GP respectively at 7-10% loading.[12] A quite significant enhancement of tensile strength and Young modulus of 25 and 18 % respectively was achieved for PLA containing 0.2 wt.% of lyophilized rGO.[4] Bao et al. reported 35% of tensile strength improvement for the composite containing an optimum amount of graphene of 0.08wt.%.[7] An enhancement of tensile strength from 36.64 to 51.14 MPa was also measured for the composite prepared by chloroform blending and containing 2 wt.% of graphene obtained via the exfoliation of graphite in chloroform.[13] As mentioned above, the use of various types of GMB and different techniques of composites preparation can lead to different results. The reduction of tensile strength, due to the interfacial defects, was observed in 3D-printed PLA filament after addition of graphene and carbon nanofibers.[14] Nevertheless the strengh and toughness changed with orientation of the layer with possible enhancement of toughness.[14] On the other hand, positive mechanical reponses were achieved with 3D-printed PLA-based graphene composites manufactured by fused filament fabrication, also orientation dependent.[15] The mechanical properties of PLA were improved also in significant manner in the composite prepared via polycondensation method, where lactic acid was functionalized with GO as crossliner for PLA chains.[10] Modifications of mechanical, barrier, thermal and microstructure properties were achieved due to the hot pressing of PLA containing graphene nanoplatelets and PEG composites.[16]

Despite several reports on the thermomechanical properties of PLA-graphene materials and, on the other hand, on functionalized graphene in view of medical applications (prosthesis, drug delivery, etc) only few examples among them deal with graphene functionalized with large bio-compatible molecules [3]. Yet, to be able to address successfully the growing demand for performant composites at long term, the synthesis of GBM should follow the sustainability rules.

Here we investigate the mechanical and thermal properties of PLA composites films containing FLG functionalized with a natural large bio-compatible system: bovine serum albumin (BSA). The composites and films are prepared via solution mixing and casting method respectively, while FLG is prepared beforehand by an efficient, simple and eco-friendly synthesis based on the exfoliation of expanded graphite in water in the presence of BSA. [17, 18]

Four series of composites are prepared: a series with a very low amount of highly dispersed FLG, two series with lyophilized FLG and last, a co-polymer composite containing plasticizer (PEG), (PEG-PLLA). The addition of plasticizers like PEG generally improves the ductility of the polymers but has contrary effect on tensile modulus and strength except if fillers are used: then all these features can be improved. [19, 20]

## 2. Results and discussion

### 2.1 Preparation and morphology

In order to evaluate the potential use of FLG obtained by simple and sustainable synthesis, [171] as well as the effect of FLG dispersion on final properties of PLA composites, four distinct series of the films were prepared. All series, series A, B, C and one series PEG-PLLA are obtained by co-mixing of two suspensions (in chloroform), suspension of polymer and suspension of FLG. Series A are composites prepared with FLG dried *via* "conventional" technique followed by dispersion in chloroform and "natural" separation step: i.e. decantation for 20h. Such preparation results in a very well dispersed and stable suspension of FLG but is characterized by a very small concentration of FLG (around 0.07 mg/mL) and high relative BSA content (c.a. 70-80%) (**Figure S1:** TGA of the FLG-BSA suspension). Series A include three FLG loadings: PLA-FLG (0.02%/A), PLA-FLG (0.03%/A) and PLA-FLG (0.07%/A). Series B and C contain the FLG dried by a lyophilization method followed by dispersion of the appropriate weight of FLG in chloroform. In series C, however, the size of the FLG flakes was initially reduced during its synthesis. It was previously shown that the use of harsher conditions, for instance longer sonication duration, leads to the reduction of FLG flakes' size.[17, 18] The preparation with lyophilization drying allowed to easily produce composites with higher FLG concentration compared to series A. Series B and C are denominated as PLA-FLG (0.17%/B), PLA-FLG (0.75%/B), PLA-FLG (3%/B); PLA-FLG (0.17%/C) and PLA-FLG (0.75%/C), according to the wt. % of FLG loading. The last series (one composite) was designed to combine two additives: FLG and a plasticizer. The PEG-PLLA-FLG (0.07%) material was prepared as in series A using a newly obtained PEG-PLLA copolymer. The optical photo of series A, C and PEG-PLLA-FLG are presented in **Figure 1**. First of all, it is worthy to note that the optical aspect of the two sides of the films differs in all series, the side in contact with Petri dish being opaque and the upper side being shinier. The same phenomenon was previously observed in the films prepared with another polymer (PVA) containing the same type of FLG.[21] These easily distinguished optical aspects can be probably related to the differences in the surface crystallinity as described below.[22] Relatively low crystallization degree was determined especially for PLA composites (table. 1). Such differences in opacity were attributed in the literature to the size of spherulites; small spherulites in less crystallized films produce weaker light diffraction, while larger spherulites in more crystallized films scatter the light with lower wavelengths giving the film an opaque appearance. [22] Different size and assembly degree of spherulites were observed in the present composites going from simple lamella *via* hedrite and sheaf to spherulite (**Figure S2**), [23] while it was challenging to distinguish the crystals from the two sides of the films. We could get a better understanding through reflexion optical micrographs of the composites described later, where the roughness of the surface can play a significant role.

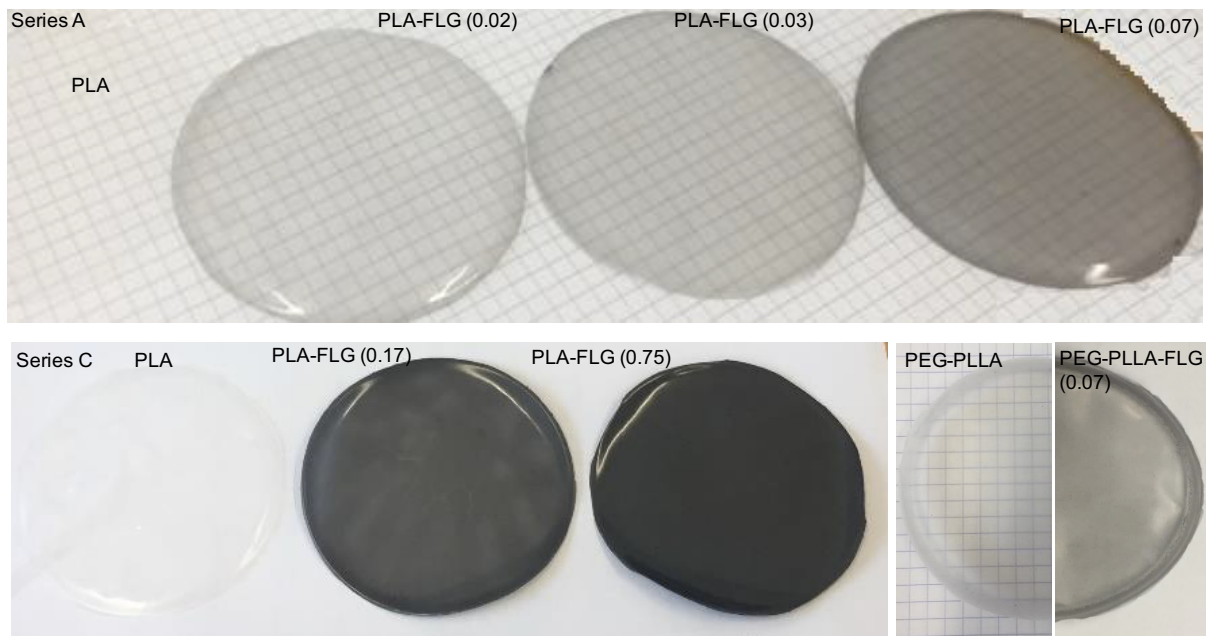

**Figure 1.** Optical images of PLA composites (series A and C), PEG-PLLA copolymer and its composite.

Dispersion, distribution and compatibility of FLG in PLA were investigated by SEM and TEM microscopies. **Figure 2** and **3** show representative SEM micrographs obtained for the composite PLA-FLG (0.07%/A), PLA-FLG (0.75%/B) and one additional composite made for conductivity measurements purpose, PLA-FLG (3%/B). Figure 2 illustrates a transversal cut, and under a certain angle, of the samples while Figure 3 displays the views of the surfaces. The observations confirm the homogenous distribution of FLG flakes within the polymer, Fig. 2 a, b and Fig. 3 c, d. We can also observe a horizontal alignment (parallel orientation) of the flakes in all composites, especially via lower resolution micrographs, Fig. 2 a, b and 3d, as well as in local analysis of the composites' surface in Fig. 3 a. According to the microscopy analysis the dispersion and compatibility of FLG into PLA is somehow dependent on the FLG content and its preparation mode. In the composite with higher FLG content, i.e. 0.75%, we could find punctually a good interface between FLG and PLA (fig. 2c inset) but relatively weak interface could be localized elsewhere, which was confirmed by TEM analysis. In Fig. 3b two FLG flakes can be seen but only one of them has a surface covered by the polymer. SEM and TEM observations suggest some aggregation of FLG *via* inter-flakes stacking that could occur either during the composite preparation or remain after dispersion of the lyophilized FLG flakes in chloroform, or both. On the contrary, a very high compatibility between FLG and the polymer can be seen in series A, with a much lower FLG content, where the suspension of FLG was initially submitted to separation and stabilization by decantation step. The transversal and surface view taken with different SEM modes, Fig. 2 b, d and Fig 3 a, shows FLG flakes that are perfectly covered by the polymer and/or BSA. The high BSA content in FLG-BSA (80%) in series A and D can indeed play a role in the dispersion and interactions with the polymer. The SEM micrographs of FLG after 20h of decantation in chloroform, where FLG is highly covered by BSA, are presented in **Figure S3**. In agreement with PLA composites, the SEM analysis of PEG-PLLA-FLG film morphology shows FLG alignment and relatively high distribution of FLG as it is a case of series A, **Figure S4**.

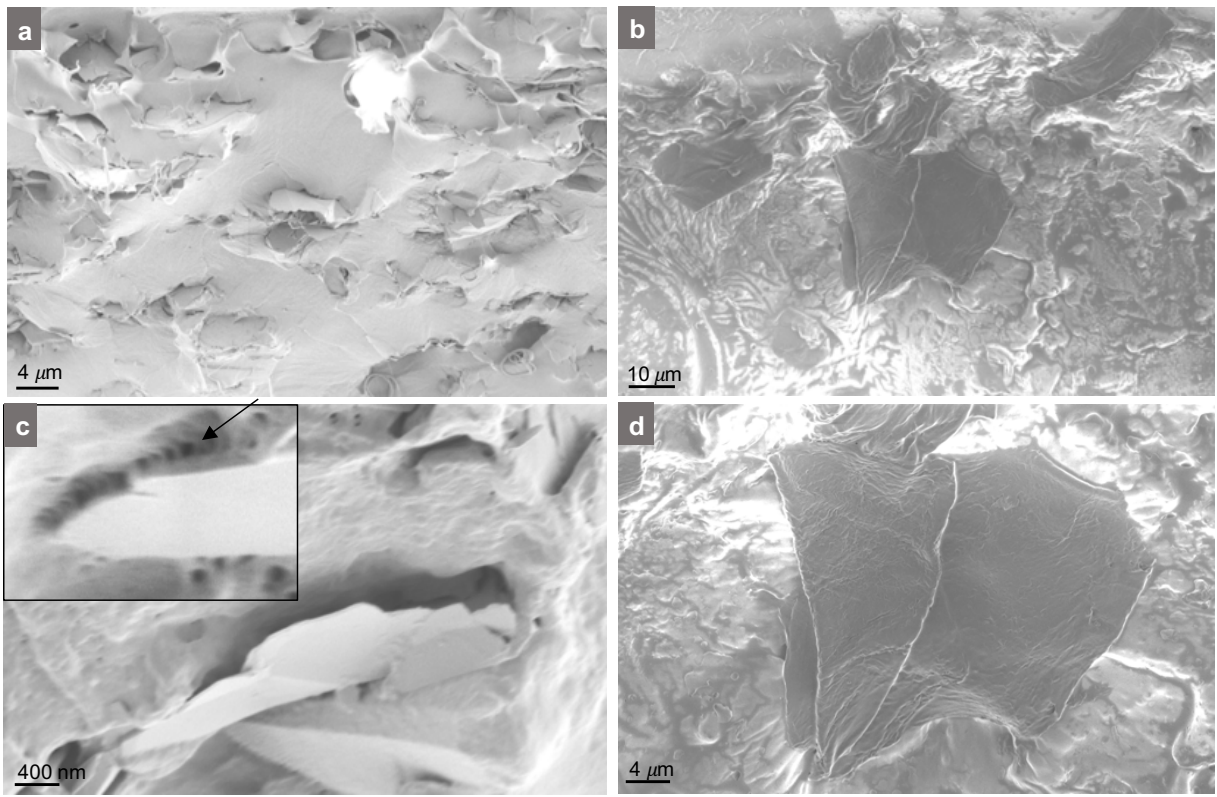

**Figure 2.** Representative SEM micrographs, transversal cut, of a, c) PLA-FLG (0.75%/B) and b, d) PLA-FLG (0.07%/A).

The presented TEM micrograph, fig. 3b, also confirms the presence of few sheets within the flakes, the two flakes containing 6 and 9 sheets respectively (edges counting). As mentioned above, an additional PLA-FLG composite film with 3% wt. of FLG was prepared since no charge transport was measured at lower loading (neither at 0.75% wt. nor at 2% wt.). This confirms that, despite homogenous FLG distribution, an additional planar stacking of FLG (in the z direction) and/or relatively low interface between FLG and PLA occur.

In agreement with the conductivity measurements results we can attribute now the FLG stacking and its consequently lower dispersion degree in chloroform to the stacking occurring during the FLG drying process. In previously obtained PVA-FLG composites, where the drying step of FLG could be omitted, the conductivity was measured already at 0.5% wt. FLG.[21] In addition, only one side of PLA-FLG (3%/B) film is conductive, the opaque one, like in a case of PVA-FLG films.

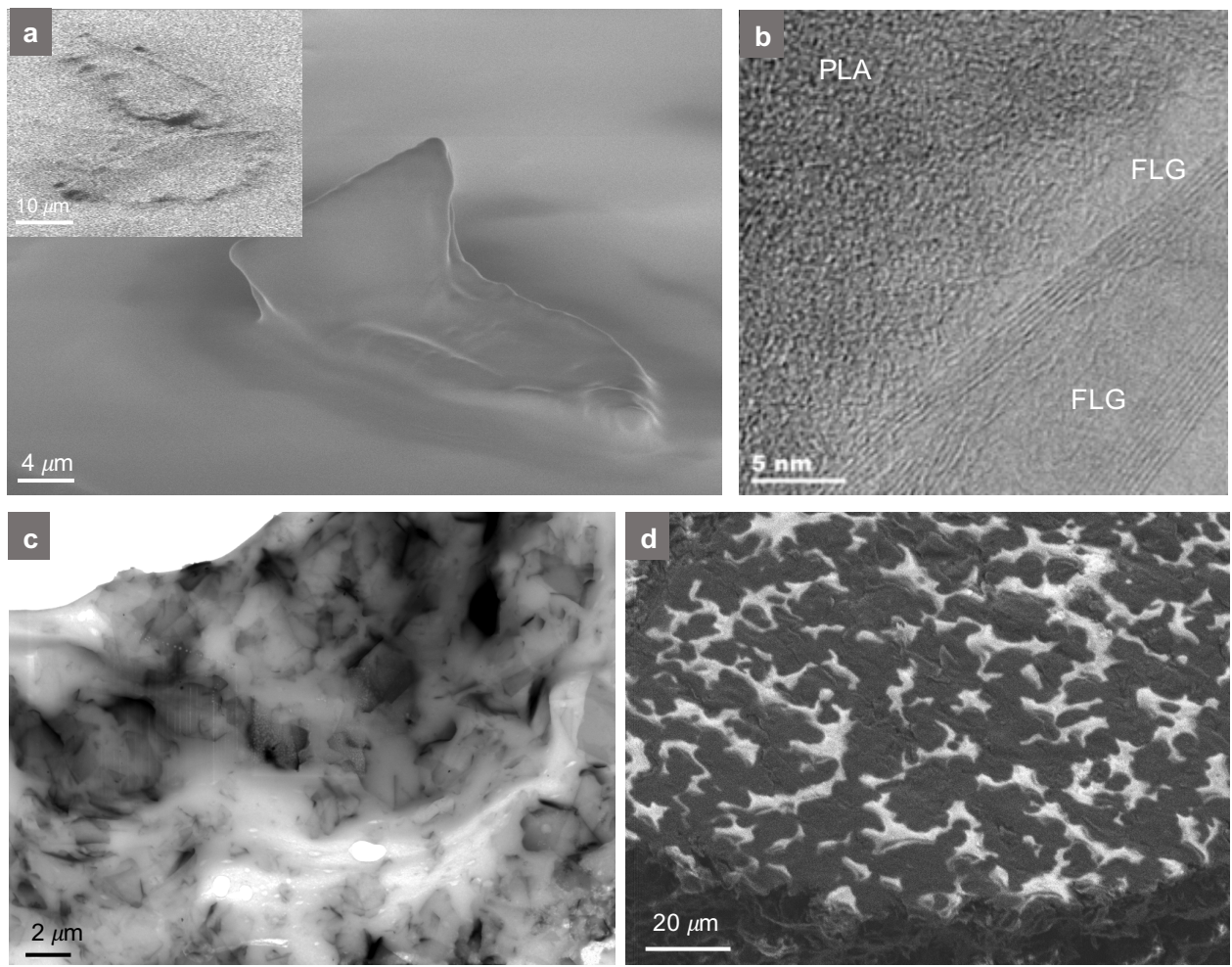

**Figure 3.** a) SEM micrograph of PLA-FLG (0.07%/A) surface, b) TEM micrograph of PLA-FLG (0.75%/B), c) STEM of PLA-FLG (0.75%/B), d) high electron intensity (6 kV) SEM of PLA-FLG (3%/B).

This phenomenon can be related to the different morphology of the two sides with easily distinguished optical aspect of both surfaces and to different FLG flakes arrangement. The optical micrographs, **Figure 4**, confirm a different topography of surfaces from both sides of the films. A much flatter morphology of the conductive surface, including a flat arrangement of the FLG flakes, can be observed in Fig. 4 (a-c vs. e-g.) as well as in the SEM micrograph, Fig. 3d. The flat arrangement of the flakes eases the contact between them (detailed studies of charge transport properties are not the goal of the present work and will be performed in further investigations for the composites with optimized FLG dispersion). Additionally, the roughness profiles obtained from the low resolution images are presented in Fig. 4d and h. The average roughness $R_a$ increases from 95 for conductive to 121 nm for non-conductive site. The rough topography of the upper site of the film is already observed at very low FLG content, i.e. 0.17% (series B), Fig. 4 i vs. k: the latter being probably the results of bubbles escaping during the drying process. The $R_a$ values for this film are 24 and 40 nm for flat and rough surface, respectively. One can see as well that $R_a$ generally increases with a higher FLG content. The XPS analysis of the surface also confirmed that the conductivity issue on one side is not due to the FLG content as the content is equivalent on both sides (c.a. 10 nm in depth). The optical micrographs show a "disturbed" flat arrangement of the flakes at the upper side due to the presence of larger polymer domains that inhibit charge transport. The measured conductivity, $5 \times 10^{-3}$ S/cm, is higher or comparable with the values reported for PLA composites containing rGO or GP additives (at the same filler amount) [24-26] and slightly higher than the one obtained previously for PVA-FLG (3%).[21]

According to the optical micrographs, a different diffraction of the light wavelenghs can be seen depending on the surface roughness (side) within the same film. The FLG flakes with flatter orientation indicates a more crystallized, and under different angle, surface exposed to interact with the light (crystals of FLG).

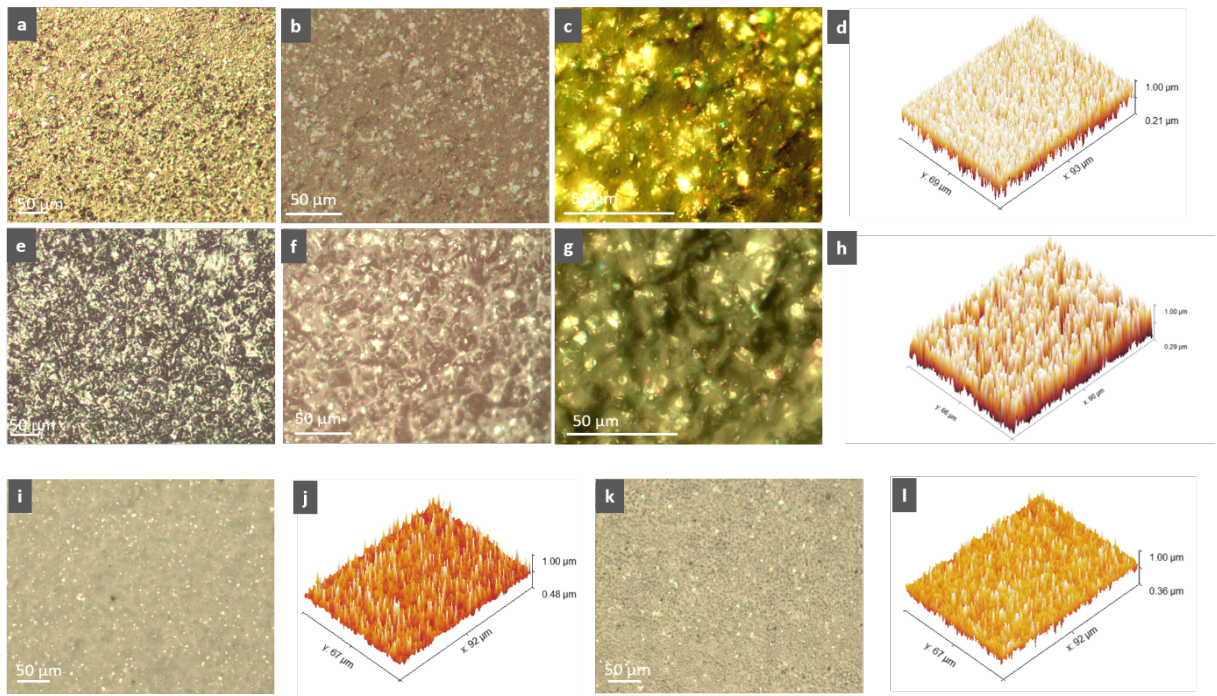

**Figure 4.** Optical micrographs of two sides of PLA-FLG film (3%/B): a-c) conductive site and e-g) non-conductive site, and i, k) of PLA-FLG film (0.17 %/B). The roughness profiles of the surface of PLA-FLG film (3%/B): d) conductive site, h) non-conductive site; of PLA-FLG film (0.17 %/B): j) and l).

## 2.2 Thermal properties

Thermal properties of the composites were investigated by TGA and DSC analysis. Selected representative curves are presented in **Figure 5**, while the values of degradation temperature ($T_d$), glass transition ($T_g$), melting temperature ($T_m$), and cold crystallization temperature ($T_{cc}$) are inserted in table 1. The table also includes the determined transition enthalpies and calculated crystallinity degrees ($\Delta H_c$, $\Delta H_{cc}$, $\Delta H_m$, %$X_c$, %$X_{cc}$, %$X_m$).

The investigations of thermal stability ($T_d$) reveal that most of the composites degrade slightly faster than the reference samples. Only two composites demonstrate enhanced thermal resistance: one with low FLG loading in series A (0.03 wt. %) showing an increase of $T_d$ by 6

°C and the one with the lowest FLG loading in series with reduced lateral FLG size, series C (0.17 wt. %) where $T_d$ is up shifted by 2°C. This suggests that in these two samples the compatibility between the polymer and FLG and/or dispersion of FLG are superior to other samples. A relatively low FLG charge and a smaller size of FLG flakes are expected to favour higher dispersion and distribution of FLG, and consequently higher thermal stability. Likewise, considering the high lateral size/thickness ratio of FLG (up to 10 µm/ max. to 10 graphene layers) compared to, for instance, commonly investigated rGO, a decrease of degradation temperature in most of the samples is not surprising. Even minor agglomeration of FLG will indeed induce the formation of "hot spots" propagating the heat very fast due to the high thermal conductivity of FLG.

The $T_g$ generally increases within the series with the increase of FLG amount, in line with a reduction of the PLA chains flexibility/mobility. Also, the PLA composites have a lower $T_g$ than pure PLA (by 3°C for 0.03%/A and 0.17%/C) suggesting the formation of shorter PLA chains. Only the incorporation of 3% wt. % FLG (series B) induces a higher $T_g$ than that of PLA. The $T_g$ also increases in PEG-PLLA-FLG composite vs. PEG-PLLA, despite a low FLG content in the former suggesting a limitation of PLA chains mobility.

The $T_m$ of the composites is higher than that of PLA by 2-3°C (1°C for the lowest loading of FLG, 0.02%/A). The enhancement of $T_m$ indicates the reduction of large and stable PLA nucleus and/or their branching due to the addition of FLG. Likewise, large FLG flakes would inhibit the flexibility of the chains. In the co-polymer PEG-PLLA, a bimodal $T_m$ is observed which can be attributed to the presence of thinner lamella. Here, with the addition of FLG, $T_m$ only slightly shifts, while $T_m$ of one sub-peak even slightly decreases.

The PLA and all its composites exhibit significant cold crystallization which can be observed by an exothermic peak between 80 and 140°C, between $T_g$ and $T_m$, Figure 5c, d. The corresponding enthalpy of cold crystallization ($\Delta H_{cc}$) as well as the enthalpy of melting ($\Delta H_m$) from the second heating run, and related crystallinity (%$X_{cc}$, %$X_m$) values are compiled in table

1. The differences between %$X_{cc}$ and %$X_m$ are very small confirming that crystallization occurs entirely during the heating. In general, the crystallization significantly decreases in the composites compared to pure PLA. With the addition of FLG the $T_{cc}$ increases of c.a. 3-6 °C indicating a limitation of PLA chains mobility and consequently of the growth of crystals. A similar behavior was reported for instance for some PLA-GO composites, while for other PLA-GO composites, on the contrary, a nucleation effect with the addition of graphene was reported.[21] These discrepancies were attributed to the differences in PLA structure (stereostructures) as well as the nature of GO additives i.e. size, dispersion.[27] In the present work the $T_{cc}$ increase is more pronounced at higher FLG loading within all series, which is probably related to adequately decreasing FLG dispersion.

None of the PLA samples crystallize during the cooling process which can be however influenced somehow also by the cooling rate and their Tcc (on heting samples) are located from 121.7 to 127.4 °C [28, 29] On the contrary, the PEG-PLLA and PEG-PLLA-FLG (0.07%) show crystallization peaks appearing during the cooling at 106.4 and 109.3 °C respectively, Figure 5 b. The crystallization degree (table 1) decreases only slightly in the composite with PEG compared to the FLG-free co-polymer.

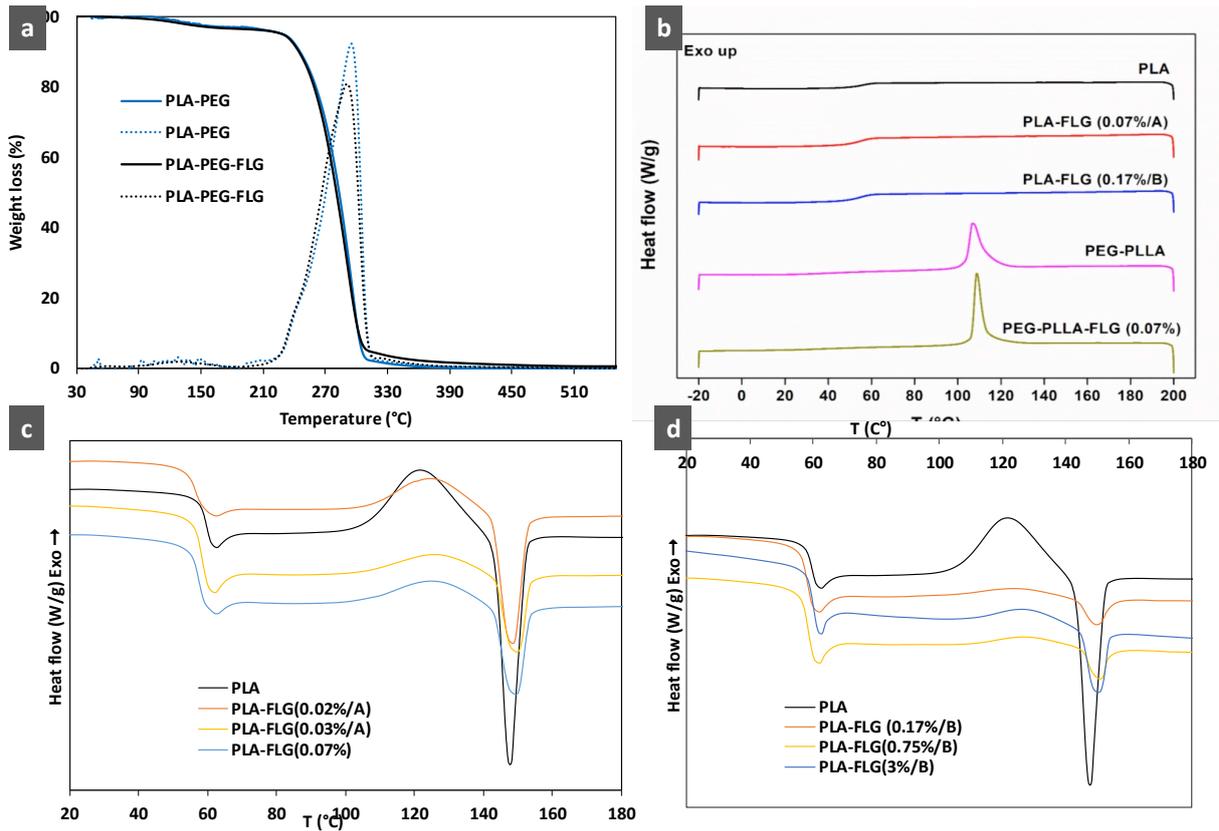

**Figure 5.** Representative, selected TGA and DSC curves: a) TGA curves of PEG-PLLA and PEG-PLLA -FLG, b) DSC curves of selected samples, c and d) DSC curves of series A and B.

Concerning the PLA composites, the reduction of crystallization is more significant for the sample more charged with lyophilized FLG, i.e. series B and C. This can be related to a lower dispersion and consequently an overall lower compatibility between FLG and PLA compared to series A, in agreement with microscopy observations. For the same reason the crystallinity also decreases with increasing of FLG charge within these series.

The differences in crystallinity between PLA series and PEG-PLLA films are also eye-visible; well-crystallized PEG-PLLA and PEG-PLLA -FLG (0.07%) films have more opaque aspect contrary to the lucent, shinier PLA series, even when compared to the more opaque side of PLA films.

**Table 1.** Temperature degradation ($T_d$), glass transition ($T_g$), melting temperature ($T_m$), and cold crystallization temperature ($T_{cc}$), determined transition enthalpies and calculated crystallinity degree ($\Delta H_c$, $\Delta H_{cc}$, $\Delta H_m$, %$X_c$, %$X_{cc}$, %$X_m$) for reference and all composites.

| Sample | $T_d$ (max-DTG) | $T_g$ (°C) | $T_{cc}$ (°C) | $T_m$ (°C) | $\Delta H_c$ (J/g) | $\Delta H_{cc}$ (J/g) | $\Delta H_m$ (J/g) | %$X_c$ | %$X_{cc}$ | %$X_m$ |
|---|---|---|---|---|---|---|---|---|---|---|
| PLA | 384.5 | 59.7 | 121.7 | 147.7 | - | -16.3 | 15.3 | - | 17.5 | 16.4 |
| PLA-FLG (0.02%/A) | 360.7 | 56.4 | 124.8 | 148.5 | - | - 7.9 | 10.0 | - | 5.5 | 8.5 |
| PLA-FLG (0.03%/A) | 390.4 | 58.4 | 126.8 | 149.7 | - | - 5.1 | 4.2 | - | 5.5 | 4.5 |
| PLA-FLG (0.07%/A) | 378.1 | 57.2 | 125.2 | 149.4 | - | - 5.0 | 5.3 | - | 5.4 | 5.7 |
| PLA-FLG (0.17%/B) | 380.2 | 58.0 | 124.1 | 149.9 | - | - 2.4 | 1.9 | - | 2.6 | 2.1 |
| PLA-FLG (0.75%/B) | 370.2 | 58.3 | 127.4 | 150.7 | - | - 2.0 | 1.8 | - | 2.2 | 1.9 |
| PLA-FLG (3%/B) | 367.0 | 60.3 | 127.1 | 150.3 | - | - 3.7 | 4.1 | - | 4.0 | 4.5 |
| PLA-FLG (0.17%/C) | **386.5** | 56.8 | 125.4 | 149.6 | - | - 1.8 | 1.7 | - | 1.9 | 1.9 |
| PLA-FLG (0.75%/C) | 365.4 | 59.0 | 126.2 | 150.4 | - | - 3.5 | 3.2 | - | 3.8 | 3.5 |
| PEG-PLLA | 295.9 | 41.7 | - | 166./160.2 | 36.2 | - | 40.8 | 38.6 | - | 43.8 |
| PEG-PLLA-FLG (0.07%) | 292.3 | 42.7 | - | 166.3/160.5 | 35.8 | - | 36.6 | 38.2 | - | 39.3 |

## 2.3 Mechanical properties

The tensile modulus, tensile strength and elongation at break for all the composites compared to reference samples, PLA and PEG-PLLA, are shown in table 2. The representative tensile-strain curves are presented in **Figure 6**. Except for the most charged samples in series B (PLA-FLG (0.75%/B and 3%/B)), the addition of FLG induces an increase of the mechanical properties. The enhancements of tensile modulus and strength in series B and C, and especially in PLA-FLG (0.17%/B), are exceptional, table 2. Such high improvement can be first related to the horizontal alignment (parallel orientation) of the FLG flakes within the polymer matrix as observed by SEM microscopy, fig. 2 a, and well-crystallized lattice of the flakes. It was indeed shown in previous studies (see introduction) that the lack of such alignment in PLA composites prevented any significant improvement.[6] Concerning the films from series A and co-polymer, with well dispersed and low % of FLG, a higher ductility compared to series B and C can be measured. Since the increase of tensile modulus and strength are less blatant, the elongation at break in series A is the most significant within all PLA series. Still, considering such low loading of FLG as 0.02% wt. in series A, the enhancement of tensile modulus and strength is very high. To exclude the impact of BSA on mechanical properties in series A (in which the percentage of BSA is significant), PLA films containing 0.08% of BSA were tested. No significant modification of elongation at break was measured for PLA-BSA films (**Figure S5**). It should be underlined that, despite an identical preparation procedure that the one applied for series A and D (i.e. sonication of BSA in water and chloroform), the structure of BSA is modified, since probably influenced by the adsorption on/interactions with FLG surface.[30, 31] We previously observed the reticular structure of adsorbed BSA over FLG after sonication in water.[21] BSA has no effect on the mechanical properties in PLA-BSA (at least at such low amount), but its possible impact on FLG dispersion in the composites cannot be excluded. Concerning series B and C it seems that the optimum FLG loading is 0.17%, then at 0.75% wt. the enhancement of all mechanical properties is less important and elongation at break is even lower than for PLA. This lower improvement is more pronounced at 3% FLG (PLA-FLG (3%/B). The increase of tensile modulus and strength is more significant in series with higher lateral size FLG, i.e. in series B, compared to series C even when comparing different FLG content. PLA-FLG (0.75%/B) shows higher enhancement than PLA-FLG (0.17%/C). This agrees with probable interactions between PLA and FLG surface and edges. We suggest that the interactions between FLG surface and PLA are more efficient, while interactions between FLG edges and PLA, if any, have much less impact on tensile modulus and strength. In FLG flakes with lower lateral size (series C) the ratio of edges-to-surface is indeed more important bringing also different chemistry. Since PLA is a hydrophobic polymer, its interactions with FLG flakes will occur mostly via hydrophobic FLG surface. It is also difficult to estimate the impact of BSA on the overall interactions either between FLG surface or edges and PLA chains. When prepared in water, the BSA bio-surfactant adsorbed on FLG surface exposes surely hydrophilic groups. After sonication/sedimentation in chloroform however the structure of BSA changes as observed by TGA and SEM analysis (Figure S1 and S3).

Compared to the PLA composites, the tensile modulus and strength are improved moderately in the co-polymer composite but the relative increase of elongation at break is notable and higher than in PLA series. This is first related to the much higher crystallinity of the co-polymer and its composite compared to the PLA series as well as to a very low initial ductility of PEG-PLLA. Likewise, the presence of PEG in the co-polymer can influence the interactions with the FLG material.

Considering that FLG flakes have a tendency to be horizontally aligned in polymers (direction of elongation), as confirmed by SEM analysis, such significant enhancement of elongation at break can indicate that interactions between FLG and PEG-PLLA essentially occur via FLG edges. This is in accordance with the chemistry of the edges (covered or not by BSA). The more hydrophilic character of the edges facilitates interactions with hydrophilic PEG groups by hydrogen bonding interactions.

**Table 2.** Mechanical properties and their enhancement (%) of PLA, PEG-PLA and their composites containing FLG.

| Sample | Tensile modulus (MPa) | Tensile strength (MPa) | Elongation at break (%) | Increase of Tensile modulus (GPa) (%) | Increase of Tensile strength (MPa) (%) | Increase of Elongation at break (%) (%) |
|---|---|---|---|---|---|---|
| PLA | 538 | 6.7 | 130 | - | - | - |
| PLA-FLG (0.02%/A) | 879 | 12 | 207 | 63 | 79 | 59 |
| PLA-FLG (0.03%/A) | 694 | 11 | 212 | 29 | **64** | **63** |
| PLA-FLG (0.07%/A) | 589 | 14.6 | 234 | 9.5 | **118** | **80** |
| PLA-FLG (0.17%/B) | 2119 | 31 | 206 | **294** | 363 | 58 |
| PLA-FLG (0.75%/B) | 1883 | 32 | 40 | **250** | 378 | -70 |
| PLA-FLG (3%/B) | 862 | 17 | 25 | 60 | 154 | -81 |
| PLA-FLG (0.17%/C) | 1568 | 25 | 140 | **190** | **273** | 8 |
| PLA-FLG (0.75%/C) | 1735 | 26.6 | 131 | **222** | **297** | 0 |
| PEG-PLLA | 2200 | 34.3 | 8 | - | - | - |
| PEG-PLLA -FLG (0.07%) | 2315 | 43.6 | 15 | 5 | 27 | **88** |

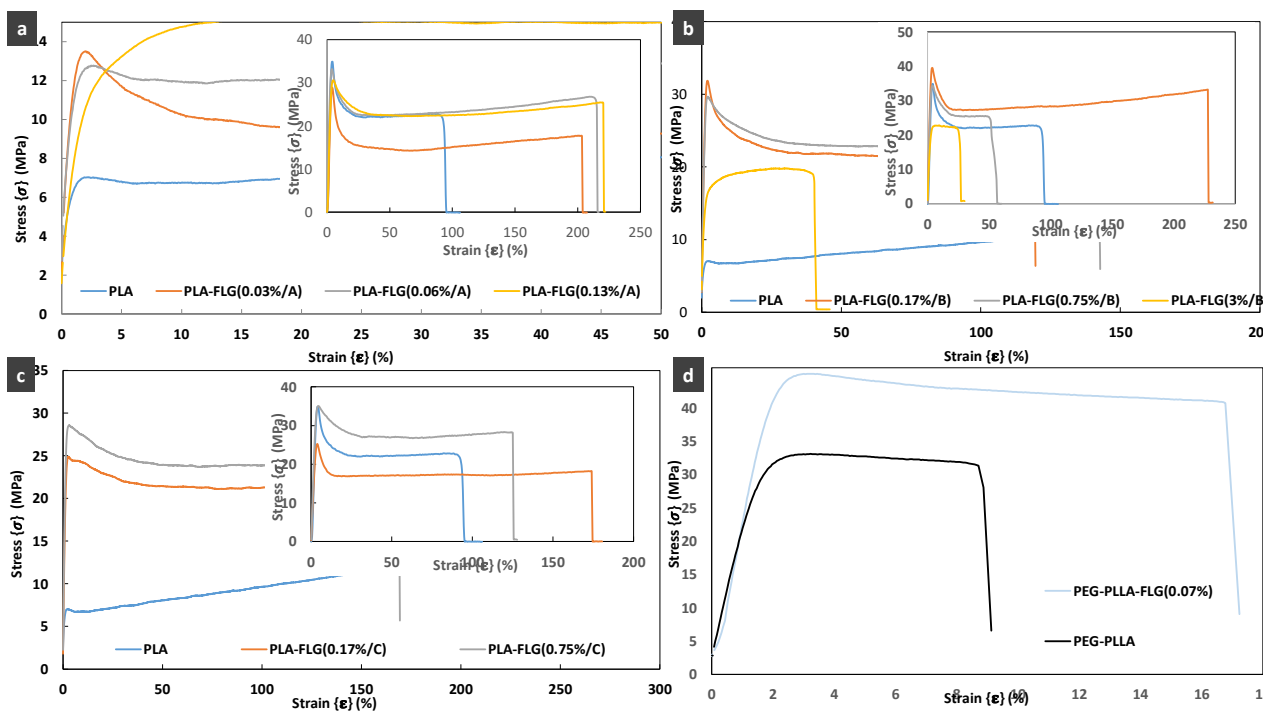

**Figure 6.** Tensile-strain curves of series A, B, C and co-polymer composite.

Likewise, the ductility behaviour remains related to the crystallinity degree of the composites with low FLG loading, i.e. series A, PEG-PLLA composite as well as series B and C for the 0.17% wt. FLG. The ductility increases with a decrease of crystallinity. At higher FLG loading this relation is however less evident and despite low crystallinity degree the elongation at break decreases.

## 3. Conclusion

Four different series of PLA (and PEG-PLLA) composites films containing FLG were synthesized by solvent co-mixing and casting method, where the used FLG was previously prepared by rapid and efficient eco-friendly exfoliation of graphite in water. The mechanical properties, thermal stability and electrical conductivity of the composites together with their structures were primarily investigated. A great improvement of tensile modulus and strength of the films up to 300 and 400% and 200 and 300% was measured at FLG loading of 0.17 and 0.75 % wt. for the composites containing large and lower size FLG flakes, respectively. A manifold elongation at break enhancement was also measured at very low FLG loading with the highest one for the composites based on PEG-PLLA co-polymer (almost 90% for 0.07% FLG). A one-side and relatively high electrical conductivity was measured for PLA films containing 3% of FLG (5 x 10$^{-3}$ S/cm). The improvement of mechanical properties in PLA-FLG films is attributed to the intrinsic properties of large size and well crystalized FLG, to their relatively good interactions with the polymer (also *via* adsorbed albumin BSA), and especially, to their parallel, horizontal orientation. The latter is an important factor, and indeed difficult to achieve with e.g. easily wrinkled GO (rGO) or other graphene platelets material as previously reported.

## 4. Experimental Section/Methods

Few layer graphene (FLG) was prepared according to the method described earlier. [17, 18] In brief, the method is based on the exfoliation of expanded graphite (EG) in water in the presence of bovine serum albumin (BSA) as bio-surfactant under stirring-assisted ultrasonication for 2 or 5 hours.

Polylactide (PLA) with molecular weight of 150 kDa was purchased from NatureWorks@ (PLA 4043D).

Polylactide-polyethyleneglycol co-polymer (PEG-PLLA) was prepared *via* ring-opening polymerization of *L*-lactide (555 equiv) initiated by a OMe-mono-capped PEG-OH ($M_n$ = 1.9 kDa; 1 equiv) and catalyzed by 1 equiv of Zn(C$_6$F$_5$)$_2$ (Scheme 1).[32, 33] Zn(C$_6$F$_5$)$_2$ was prepared according to a literature procedure. [34] Co-polymer was analyzed via $^1$H NMR and GPC (**Figure S6** and **S7**). GPC data agree with a mono-disperse material with a narrow polydispersity ($M_n$ = 80600 g/mol, c.a. 78700 g/mol (PLLA) and 1900 g/mol for PEG). The obtained $M_n$ value was multiplied by the corrective factor of 0.58 for an estimation of the real PLA Mn value. [35] More details of PEG-PPLA preparation are in SI.

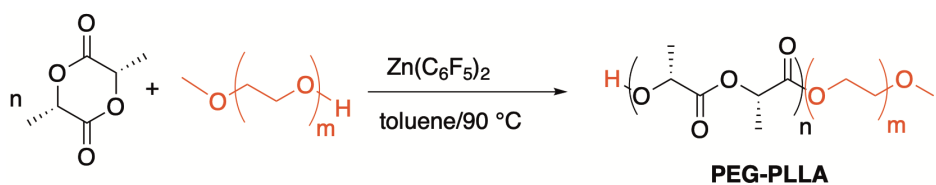

Scheme 1. Ring-opening polymerization of L-lactide in the presence of PEG-OH catalyzed by $Zn(C_6F_5)_2$ for the production of PEG-PLLA.

Chloroform was purchased by Carlo Erba Reagents.

Synthesis of the films:

PLA: PLA was dissolved in chloroform via stirring for 3h in r.t. with concentration of 72 mg/mL. The suspension was spilled on Petri dishes to be left for drying at r. t. for 20 h and next 3 days at 40°C under vacuum.

PEG-PLLA: PEG-PLLA powder was dissolved in chloroform under stirring for 3 h with the concentration of 27 mg/mL. The suspension was spilled on Petri dishes to be left for drying at r. t. for 20 h and next 3 days at 40°C under vacuum.

Series A: PLA was dissolved in chloroform via stirring for 3h in r.t. with concentration of 72 mg/mL. Separately, FLG was dispersed in chloroform via ultrasonication at the concentration of 0.25 mg/mL for 1h and next the suspension was left for decantation step for 20 h. The appropriate volume of the supernatant with concentration c. a. 0.07 mg/mL was then separated to be added to the PLA suspension. All was mixed for next 0.5 h and spilled on Petri dishes to be left for drying at r. t. for 20 h and next 3 days at 40°C under vacuum.

Series B and C: PLA was dissolved in chloroform via stirring for 3h in r.t. with concentration of 84 mg/mL. Separately, FLG obtained as aqua colloid after 2h of ultrasonication (series B) or 5h of ultrasonication (series C) of EG was submitted to lyophylization process. The appropriate amount of lyophilized FLG was then added to the PLA suspension. All was mixed for 1.5h and mild-sonicated for 20 min, and next spilled on Petri dishes to be left for drying at r. t. for 20 h and next 3 days at 40°C under vacuum.

Series D: PEG-PLLA powder was dissolved in chloroform under stirring for 3 h with the concentration of 27 mg/mL. Separately, FLG was dispersed in chloroform via ultrasonication at the concentration of 0.25 mg/mL for 1h and next the suspension was left for decantation step for 20 h. The appropriate volume of the supernatant with concentration of c. a. 0.07 mg/mL was then separated to be added to the PEG-PLLA suspension. All was mixed for next 0.5 h and spilled on Petri dishes to be left for drying at r. t. for 20 h and next 3 days at 40°C under vacuum.

Characterization tools:

Thermogravimetric analysis (TGA) was carried out with TGA Q5000 V3.17 Build 265 under air flow with the heating rate of 50°/min. // Differential scanning calorimetry (DSC) was performed on Q200 V24.11 Build 124. Two cycles were run under $N_2$ with heating and cooling rate of 10°/min and 5°C/min respectively, (isothermal steps of 2 min). // Scanning electron microscopy (SEM) was performed on Jeol JSM-6700 F working at different accelerated voltage. For the cross-sectional analysis the films were broken/cut after being frizzed in liquid $N_2$. // Transmission electron microscopy (TEM) was performed on 2100 F Jeol microscope using the thin slices of the films cut by razor from the frizzed film. // Tensile testing was carried out on ElectroPuls E3000 (tensile modulus and strength) and Instron 4500 (elongation at break). Prior to the measurements the films were cut into specimens with appropriate dimensions for using in serrated grips. For the reason of repeatability, several tests were performed with the same sample. // Conductivity measurements (sheet resistance) of the film was performed via four points probes method (FPPs) under $N_2$ atmosphere (glove-box) on KEITHLEY 4200-SC. // Polarized light micrographs were obtained on Leica ICC 50 HD microscope.

**Supporting Information** Preparation, H[1]-NMR spectra and GPC analysis report of PEG-PLLA, TGA: BSA composites, selected optical images, SEM micrographs of BSA, BSA-FLG samples and PEG-PLLA-FLG film, tensile-strain curves of PLA-BSA film. Supporting Information is available from the Wiley Online Library or from the author.

**Acknowledgements**
We acknowledge: Walid Baazis (TEM microscopy, IPCMS Strasbourg), Thierry Romero (SEM microscopy, ICPEES), Alain Carvalho (ICS, Strasbourg) and Céline Piras (ICPEES) for POM. This work was supported by the "pre-maturation" program CNRS.